# Data Hiding and Retrieval Using Permutation Index Method

Rahul R Upadhyay
*Mechanical Engineering Department, BTech(Final Year)*
*BBDNITM, Lucknow*
rahulrupadhyay91@gmail.com

**Abstract**

*In this paper a novel approach for matrix manipulation and indexing is proposed .Here the elements in a row of matrix are designated by numeric value called 'permutation index' followed by the elements of the row being randomised. This is done for all the rows of the matrix and in the end the set of permutation indices are put in the parent matrix and random locations depending on a pre decided scheme called passkey. This passkey is used to put back the elements of all the rows back in the correct sequence. This approach finds application in data encapsulation and hiding .*

## 1. Introduction

Sound files and Image files in electronic formats are basically stored as numeric data in matrix forms. Thus data manipulation of electronic sound and image files is nothing but matrix manipulation. Two matrices having different sequence of elements belonging to the same domain set are fundamentally different and thus represent different data. We use this very basic concept and manipulate and randomise the sequence of elements that make up the matrix, thus rendering it obscure, followed by putting back the elements in the correct sequence. This approach can be used to protect all data that are essentially in matrix forms.

Sound files are read as column matrix while Image files are read as two or three dimensional matrices depending on them being greyscale, RGB etc. As already described we randomise the sequence of matrix elements and render it obscure to an unauthorized system or person. The highly randomized sets of elements are organized using a scheme called '**passkey**' which will be present by default with a person to whom the data is supposed to be delivered.

In this method each row of the matrix is analysed and the '**permutation-index**', later described, of the entire is row found out. This is followed by placement of permutation indices in the parent matrix as per a pre decided scheme between transmitter and receiver called passkey. This process is carried out for all the rows and in the end all the permutation-indices are arranged in the parent matrix according the passkey scheme that is pre decided during shuffling and alteration of rows.

The authorized person who is meant to decrypt the haphazard set of data has the **passkey** which helps to organize the elements of the matrix to fetch meaningful information. [27][28]

## 2. Related Work

This techniques holds application in the field of data hiding and thus before analysing this work we put forth a survey of various data hiding techniques

Kuo et al's work put forth a novel method to hide and conceal data by fragmenting an image into smaller blocks and then developing the histogram for each block [1][2]. This is followed by creating minimum and maximum points so as to create space to embed and hide the data. [3][4].

In Kodos {5} work we come across a method of data hiding in audio formats[26] which the polarity of reverberations are embedded in high frequency audio signals. The embedded data is fetched by correlation of addition and subtraction of audio signals.

Data hiding in video signals was proposed in Le et al[6] which is based on individual video frames that make up the entire video sequence[7]. It is an adaptive embedding algorithm wherein residual 4x4 DCT blocks are scanned in an inverse zig zag fashion until first non zero coefficient is encountered. This coefficient is compared with a predefined value and embedding is done if the coefficient is larger.[8][9]

The authors in [10] have provided a novel approach to hide more data in an image wherein the histogram of blocks of an image are shifted to a point corresponding to the minimum point on the histogram and data is stored between these points.

A bit plane splicing algorithm was proposed by Naseem et al [11]. In this approach rather than hiding the data in pixel by pixel we store data depending on the intensity of individual pixels. Various ranges are defined for various pixels and data bits are stored randomly in the matrix rather than being adjacent to each other. Similar approach was implemented by the authors of [12][13] to hide data in halftone images.

## 3. Sequence Alteration or Shuffling.

For description purpose we take an arbitrary 5x5 matrix "A".

**A:**

|    | C1 | C2 | C3 | C4 | C5 |
|----|----|----|----|----|----|
| R1 | 17 | 24 | 1  | 8  | 15 |
| R2 | 23 | 5  | 7  | 14 | 16 |
| R3 | 4  | 6  | 13 | 20 | 22 |
| R4 | 10 | 12 | 19 | 21 | 3  |
| R5 | 11 | 18 | 25 | 2  | 9  |

Table 1. Matrix A

- Elements of row R1= {.17, 24, 1, 8, 15}
- Number of ways elements can be shuffled: 5! = **120**
- Probability of forming correct sequence of the elements in row 1= 1/120 = **P(R1)** = .008333

If all the elements of all the rows are randomized then the probability of finding the correct sequence of all the elements in all the rows :

- P(R1) x P(R2) x P(R3) x P(R4) x P(R5) = 1/5! x 1/5! x 1/5! x 1/5! x 1/5! = **4.018 x 10^-11**

It can be clearly seen that brute force attack to guess all the elements of even a small 5 cross 5 matrix is implausible owing to such minimal probability. Thus for higher order matrices brute force guessing is impractical. It is more convenient to use this approach for matrices having less than or around 8 or 9 columns in each row as higher number of columns lead to big and complex 'permutation matrices', described henceforth [14].

## 4. Permutation Index (#P)

It was earlier stated that rows of matrices are designated by a numeric value called permutation index.

Def of Permutation Index: **Permutation index of a row in a matrix, say R1, carrying n elements is the row index number of a matrix, called Permutation matrix, which represents all the permutations of the elements of the row R1 taken n at a time and in reverse lexicographic order.**

Let's take an arbitrary 3x3 matrix:

R1 8 1 6
R2 3 5 7
R3 4 9 2

Here elements of third row or R3= {4, 9, 2}
Number of ways elements of R3 can be shuffled= 3!= 6
Lets put all the possible permutations of R3 in a matrix P. Then,

**P:**

| Rows/Columns | C1 | C2 | C3 |
|--------------|----|----|----|
| 1 | 2 | 9 | 4 |
| 2 | 2 | 4 | 9 |
| 3 | 9 | 2 | 4 |
| 4 | 9 | 4 | 2 |
| 5 | 4 | 9 | 2 |
| 6 | 4 | 2 | 9 |

Table 2. Probability Matrix

A matrix containing all the permutations of a set of elements in a sequence would also be its superset. As per the above definition the permutation index of row R3 will be the row which represents all the elements of R3 in the exact sequence. Thus the permutation index of R3 here is, 5 or **#P(R3)=5**. All the rows are computed[15][30] for permutation index via looping wherein each row is computed for permutation index, one at a time.

### 4.1 Points to Note about Permutation Index Method.

- The number elements in a row to be shuffled and computed for permutation indices in each loop , termed "**column-constant**",[16][29] should be less than 10 as higher number of elements lead to mammoth permutation matrices which takes too much processing time to compute. Column constant is a pre decided factor and is confidential between the receiver and the transmitter and thus provides extra security. [21][22]

- For a M cross N matrix having "x" as its column-constant, the number of permutation indices are, O= **(M X N)/x.** These O elements are placed in the parent matrix as per the pre decided scheme called 'passkey' Thus O elements are placed in the parent matrix, thus increasing the number of elements to (MxN)+O. Higher the column-constant, lower the O. Column constant should be chosen between 2-9 such that O is a positive integer.[17]18]

- O elements are arranged in the parent matrix by adding Mx amount of rows at the end of the parent matrix according to the following relation.
Let's assume that the parent matrix has M cross N rows and columns and the column constant used is x then number of rows and columns added at the end of the parent matrix **, Mx , should be such that Mx X N >= O.** In case that Mx X N is not equal to zero, then the remaining places will be filled with any pre decided arbitrary number.

Following figure represents a 3 x 4 matrix having column constant of 2 . Thus O= (3x4)/2= 6. These extra six elements are arranged such that **Mx X 4 >=6.** The yellow boxes represent the scheme chosen for housing permutation indices.

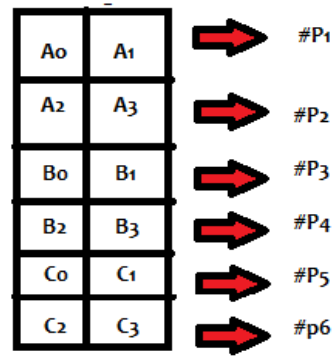

Figure 1. Placement of Permutation Indices in main.

## 5. Re organizing rows of matrix

At receiving end, the matrix is reorganised on the basis of passkey and column constant(X) as per following steps:

- The compound matrix is fragmented into two matrices, L and K such that L= MxN and K is the Mx cross N matrix at the bottom of the L matrix.[19][25]
- Now all the elements of L are distributed sequentially in a J x X matrix such that J= (MN/x).
- Now all the rows of this JxX matrix is read and a permutation matrix for all the elements in each individual row is made.
- Corresponding to each row, the permutation index is fetched from the passkey scheme and the row corresponding to the permutation index of the permutation matrix is taken and replaced with the current row of the JxX matrix.
- The JxX matrix is finally converted back to the MxN matrix and this is the desired data.

Lets yet again take a random matrix

| B | C1 | C2 |
|---|---|---|
| R1 | 1 | 3 |
| R2 | 4 | 2 |

Table 3. Matrix B

Taking column constant x=2; permutation matrices for row 1 and 2 are

| For R1 | C1 | C2 |
|---|---|---|
| 1 | 1 | 3 |
| 2 | 3 | 1 |

Table 4. Probability Matrix of Row R1

| For R2 | C1 | C2 |
|---|---|---|
| 1 | 2 | 4 |
| 2 | 4 | 2 |

Table 5. Probability Matrix of Row R2

This implies that the permutation index of row1 and row 2 are '1' and '2' respectively. Now we place the permutation indices in the parent matrix B as per the passkey scheme shown in diagram to obtain the compound matrix after randomising the elements of row1 and row 2.

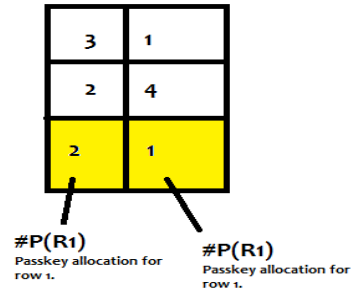

Figure 2. Placement of Permutation Indices in main.

To fetch the data back, we perform the reverse algorithm.[20] The passkey for each row is fetched from the passkey scheme :

**#P(R2)= 2**
**#P(R1)=1**

The permutation matrices of compound matrix is obtained as given in table 4 and table 5.The rows R1 and R2 of the compound matrix is replaced with rows #P(R1) and #P(R2) of the permutation matrices to obtain matrix B as given in Table 3.

## 6. Matlab source code for performing Permutation Index Calculation for rows and shuffling in a matrix.

### 6.1. Creating Permutation Matrix.

In matlab **perms (v)** is a function that returns a matrix containing all the possible permutations of a set of numbers, v in a REVERSE LEXICOGRAPHIC ORDER.

**Example**      V= [ 1 2 3 ];
**And,**         A= **perms (v)**;
**Then A=**      3 2 1
                 3 1 2
                 2 3 1
                 2 1 3
                 1 2 3
                 1 3 2

For finding the permutation matrix of a vector the vector is first sorted in ascending order and then perms() is applied.

Thus for a vector A= [ 4 2 1 3 5], the permutation matrix is obtained by the following syntax:

**perms(sort(A));**

## 6.2. Computing the Permutation index

Permutation index of a vector is the row index of the permutation matrix that corresponds to the exact sequence of the vector.

If A is the permutation matrix of [1 2 3] then the permutation index of [1 2 3 ] will be "5" .

The syntax for computing permutation index is given as[24]

**find(ismember(M,X,'rows'));**

where M is the permutation matrix and X is the row for which the permutation index has to be computed.

## 6.3. Random Shuffling of elements a vector(or row of a matrix).

If a matrix A= [ 2 4 5]; then the syntax to randomly shuffle its elements is given by :[23]

**A(randperm(length(A)));**

'randperm' is used to find random permutation of a vector.

## 6.4. PIC Calculation for 5x5 matrix

```
Command Window
>> A=magic(5)

A =

    17    24     1     8    15
    23     5     7    14    16
     4     6    13    20    22
    10    12    19    21     3
    11    18    25     2     9

>> b=[];
>> for i=1:5
row_values=A(i,:);
permutation_matrix=perms(sort(row_values));
M=permutation_matrix;
X=row_values;
b(i)=find(ismember(M,X,'rows'));
end
>> plot(b)
>> grid on
```

Figure 3. Matlab Source Code for computing permutation indices of matrix A.

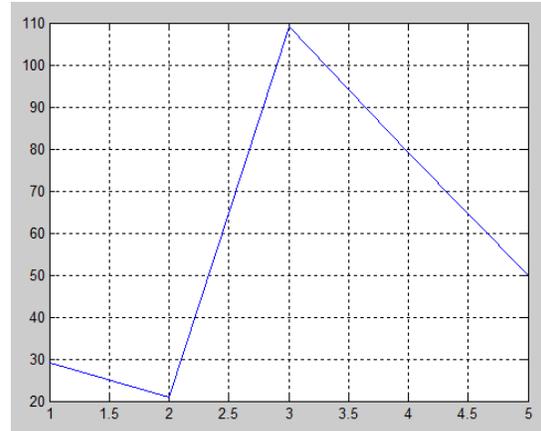

Figure 4. Plot of permutation indices of rows of matrix A.

## 7. Conclusion

This method finds application in all fields of engineering and science where data acquisition, storage or transmission of confidential numeric entities is needed.[29][31].

The biggest disadvantage of this approach is that the parent matrix whose rows are randomised to render them obscure are concatenated with passkey matrix carrying permutation indices, thus increasing the size in reverse proportionality to the column constant .

This can be circumvented by choosing a bigger column constant at the expense of processing speed.

This paper introduces the concept of permutation index which can find application in a wide range of engineering problems including steganography, data-acquisition and data compression.

## References

[1] Wen-Chung Kuo, Dong-Jin Jiang, Yu-Chih Huang, "A Reversible Data Hiding Scheme Based on Block Division", Congress on Image and Signal Processing, Vol. 1, 27-30 May 2008, pp. 365-369

[2] Yih-Chuan Lin, Tzung-Shian Li, Yao-Tang Chang, Chuen-Ching Wang, Wen-Tzu Chen, "A Subsampling and Interpolation Technique for Reversible Histogram Shift Data Hiding", Image and Signal Processing, Lecture Notes in Computer Science, Vol. 6134, 2010, Publisher: Springer Berlin/Heidelberg, pp. 384-393.

[3] Chyuan-Huei Thomas Yang, Chun-Hao Hsu, "A High Quality Reversible Data Hiding Method Using Interpolation Technique," IEEE Fifth International Conference on Information Assurance and Security, Vol. 2, 18-20 Aug. 2009, pp. 603-606.

[4] Che-Wei Lee and Wen-Hsiang Tsai, "A Lossless Data Hiding Method by Histogram Shifting Based on an Adaptive Block Division Scheme", Pattern Recognition and Machine Vision, River Publishers, Aalborg, Denmark, pp. 1–14.

[5] K. Kondo, "A Data Hiding Method for Stereo Audio Signals Using the Polarity of the Inter-Channel Decorrelator", IEEEK. Kondo, "A Data Hiding

Method for Stereo Audio Signals Using the Polarity of the Inter-Channel Decorrelator", IEEE

[6] Yu Li, He-xin Chen, Yan Zhao, "A new method of data hiding based on H.264 encoded video sequences", IEEE 10th International Conference on Signal Processing(ICSP), 24-28 Oct. 2010, pp. 1833-1836.

[7] Yu Li, He-xin Chen, Yan Zhao, "A new method of data hiding based on H.264 encoded video sequences", IEEE 10th International Conference on Signal Processing(ICSP), 24-28 Oct. 2010, pp. 1833-1836.

[8] Wojciech Mazurczyk and Krzysztof Szczypiorski, "Steganography of VoIP streams", On the Move to Meaningful Internet Systems, OTM 2008, Springer, Vol. 5332, pp. 1001-1018.

[9] Al-Sadi, A. and El-Alfy, E., (2011) "An Adaptive Steganographic Method for Color Images Based on LSB Substitution and Pixel Value Differencing", *ACC 2011, Part II, CCIS 191*, pp. 535–544.

[10] Wen-Chung Kuo, Dong-Jin Jiang, Yu-Chih Huang, "A Reversible Data Hiding Scheme Based on Block Division", Congress on Image and Signal Processing, Vol. 1, 27-30 May 2008, pp. 365-369.

[11] M. Naseem, Ibrahim M. Hussain, M. Kamran Khan, Aisha Ajmal, "An Optimum Modified Bit Plane Splicing LSB Algorithm for Secret Data Hiding", International Journal of Computer Applications, Vol. 29, No. 12, 2011. Foundation of Computer Science, New York, USA, pp. 36-43.

[12] Ming Sun Fu and O.C. Au, "Data hiding watermarking for halftone images", IEEE Transactions on Image Processing, Vol.11, No. 4, Apr. 2002, pp.477-484.

[13] Soo-Chang Pei and J.M. Guo, "Hybrid pixel-based data hiding and block-based watermarking for error-diffused halftone images", IEEE Transactions on Circuits and Systems for Video Technology, Vol.13, No. 8, Aug. 2003, pp. 867- 884.

[14] Animesh Kr Trivedi, Rishi Kapoor, Rajan Arora, Sudip Sanyal and Sugata Sanyal, RISM - Reputation Based Intrusion Detection System for Mobile Ad hoc Networks, Third International Conference on Computers and Devices for Communications, CODEC-06, pp. 234-237.

[15] Sandipan Dey, Ajith Abraham, Bijoy Bandyopadhyay and Sugata Sanyal, "Data Hiding Techniques Using Prime and Natural Numbers", Journal ofDigital Information Management, Volume 6, 2008.

[16] Nosrati Masoud, RonakKarimi, HamedNosrati,Ali Nosrati, "Taking a Brief look at steganography: Methods and Approaches", Journal of American Science, vol.7, Issue 6, 2011, pp. 84-88.

[17] C. Wang, Q. Wang, K. Ren, N. Cao and W. Lou, "Toward Secure and Dependable Storage Services in Cloud Computing, " IEEE Transactions on Services Computing, vol. no. 5, art. no. 5765928, pp. 220-232.

[18] Shital C. Patil, R. R. Keole, "Cryptography, Steganography & Network Securities", "International Journal of Pure and Applied Research in Engineering and Technology", 2012; Volume 1(8): pp. 9-15

[19] Dey, Sandipan, Ajith Abraham, and Sugata Sanyal. "An LSB Data Hiding Technique Using Natural Number Decomposition", IEEEThird International Conference onIntelligent Information Hiding and Multimedia Signal Processing, 2007. IIHMSP 2007, vol.2. 2007, pp. 473-476.

[21] Yu Li, He-xin Chen, Yan Zhao, "A new method of data hiding based on H.264 encoded video sequences", IEEE 10th International Conference on Signal Processing(ICSP), 24-28 Oct. 2010, pp. 1833-1836.

[22] R. A. Vasudevan, A. Abraham, Sugata Sanyal, D.P. Agarwal, "Jigsaw-based secure data transfer over computer networks", Int. Conference on Information Technology: Coding and Computing, pp. 2-6, vol.1, April, 2004.

[23] Description of random permutation in matlab : http://www.mathworks.in/help/matlab/ref/randperm.html

[24] Description of 'find' method in matlab : http://www.mathworks.in/help/matlab/ref/find.html

[25] Shantanu Pal, Sunirmal Khatua, Nabendu Chaki, Sugata Sanyal ,"A New Trusted and Collaborative Agent Based Approach for Ensuring Cloud Security", International Journal of Engineering; scheduled for publication in Vol. 10, Issue 1, February, 2012. ISSN: 1584-2665.

[26] Rahul R Upadhyay "Study of Encryption and Decryption of Wave File in Image Formats" International Journal of Advanced Networking and Applications.ISSN: 0975-0290. pages :1847-1852.

[27] Soumyendu Das, Subhendu Das, Bijoy Bandopadhyay, Sugata Sanyal, "Steganography and Staganalysis: Different Approaches", Int. Journal of Computers, Information Technology and Engineering (IJCITAE), vol.2, no.1, June, 2008.

[28] Rohit Bhaduria,Sugata Sugal, "Survey on Security Issues in Cloud Computing and Associated Mitigation Techniques" ,International Journal of computer applications, Vol: 47,No:18,June 2012, pp:47-66.

[29] V Goyal, V Kumar, M Singh, A Abraham, S Sanyal, "CompChall: addressing password guessing attacks", Information Technology: Coding and Computing,(ITCC) 2005. Vol. 1 P 739-744 .

[30] Rohit Bhadauria, Sugata Sanyal, "Survey on Security Issues in Cloud Computing and Associated Mitigation Techniques", Intl. Journal of Computer Applications, Vol. 47, No. 18, pp. 47-66. Published by Foundation of Computer Science, New York, USA

[31] M. R. Thakur, D. R. Khilnani, K. Gupta, S. Jain, V. Agarwal, S. Sane, S. Sanyal and P. S. Dhekne, "Detection and Prevention of Botnets and malware in an enterprise network, " International Journal of Mobile and Wireless Computing, vol. no. 5, 2012, pp.144-153.

## Author Biography

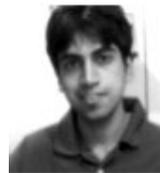

.**Rahul R Upadhyay** is a B tech student of Department of Mechanical engineering, BBD National Institute of Technology and Management. He has an interdisciplinary research interest in topics varying from embedded electronics, network security, fluid mechanics and thermodynamics. He has published papers in national as well as international journals and participated and won several technical awards in national level competitions and conferences.